\documentclass[%
 reprint,
 amsmath,amssymb,
 aps,
 prl,
]{revtex4-1}

\usepackage{graphicx}
\usepackage{dcolumn}
\usepackage{bm}

\newcommand{\ket}[1]{\,|  #1 \rangle}				

\newcommand{\xgate}{\mathsf{X}}

\newcommand{\tgate}{\mathsf{T}}
\newcommand{\rgate}{\mathsf{T}}

\newcommand{\zgate}{\mathsf{Z}}
\newcommand{\cnot}{\mathsf{CNOT}}
\newcommand{\pgate}{\mathsf{P}}

\newcommand{\hgate}{\mathsf{H}}

\input{qcircuit} 
\newcommand{\dw}[1][-1]{\ar @{--} [0,#1]}

\newcommand{\twowire}[1]{\ar@{}[d]_-{\txt{#1}\Bigg\{}}

\begin{document}

\title{Popescu-Rohrlich correlations imply\\efficient instantaneous nonlocal quantum computation}

\author{Anne Broadbent}
 \altaffiliation[Part of this research was performed while the author was affiliated with ]{IQC, University of Waterloo.}
\affiliation{%
 Department of Mathematics and Statistics, University of Ottawa\\
}%


%

\begin{abstract}
In instantaneous nonlocal quantum computation, two parties cooperate in order to perform a quantum computation on their joint inputs, while being
restricted to a \emph{single} round of simultaneous communication.  Previous results showed that instantaneous nonlocal quantum computation is possible, at the cost of
an exponential amount of prior shared entanglement (in the size of the input). Here, we show that a \emph{linear} amount of entanglement suffices,
 (in the size of the computation),
 as
long as  the parties share nonlocal correlations as given by the \emph{Popescu-Rohlich} box.
 This means that communication is not required for efficient instantaneous nonlocal quantum computation.
 Exploiting the well-known relation to position-based cryptography, our result also implies
  the  impossibility of  secure position-based cryptography against adversaries with
 non-signalling correlations.
  Furthermore, our construction establishes a quantum analogue of the classical communication complexity collapse under non-signalling correlations.
\end{abstract}

\pacs{Valid PACS appear here}
\maketitle


In two-party quantum computation, Alice and Bob 
wish to evaluate a quantum circuit $C$
on their joint inputs. Here, we consider that Alice and Bob are \emph{co-operating} players that are restricted only in the way they communicate:
they can agree ahead of time on a joint strategy (and possibly establish shared correlations or entanglement), but they are  separated before receiving their quantum inputs, and are allowed only a \emph{single} round of simultaneous communication (thus: Alice sending a message to Bob, and Bob sending a message to Alice, \emph{simultaneously}). The requirement is that at the end of this round, Alice and Bob must share the output system $\rho_{\text{out}}^{AB}= C(\rho_{\text{in}}^{AB})$. This problem is known as \emph{instantaneous nonlocal quantum computation}. Remarkably, this task is known to be achievable for any circuit
as long as the parties share an exponential (in the size of the inputs) amount of an entangled resource given as copies of the two-qubit maximally entangled state, $\frac{1}{\sqrt{2}}(\ket{00} + \ket{11})$\cite{BCF+14,BK11}.

The motivation for the study of instantaneous nonlocal quantum computation includes the  foundations of quantum physics and distributed computing; however, the
original and main motivation is in the context of  \emph{position-based cryptography}.  Here, parties use their
geographic location as a cryptographic credential. Protocols typically exploit the relativistic no-signalling principle: the idea being that a careful timing argument would then ascertain the location of the parties\cite{BC93}. Unfortunately, a no-go result is known in the the classical context\cite{CGMO09}. Due to the quantum no-cloning principle, it was originally believed that quantum protocols could escape this impossibility result\cite{KMS11,Mal10a,CFG+10arxiv,Mal10b,LL11}. However, these protocols are all broken by  entanglement-based attacks, as long as the colluding adversaries share a large enough (exponential) amount of entanglement\cite{BCF+14,BK11}
This exponential overhead in resources (in terms of entanglement and quantum memory) leads to the main open problem in this area, which is to give a protocol which can be executed efficiently  by honest players, but for which any successful attack requires an exponential amount of resources (see related work\cite{Unr14b,CL15,Spe15arXiv}).

In an apparently unrelated line of research, Popescu and Rohlich\cite{PR94} defined the nonlocal box (\emph{NLB}) as a virtual device that  achieves the CHSH conditions\cite{CHSH69} perfectly:
when Alice (Bob) uses input~$x$~($y$), the NLB produces output~$a$~($b$) such that  $a \oplus b = x \cdot y$. We note that quantum mechanics achieves this correlations with a maximum value of $\approx 85\%$\cite{Cir80}, but that the NLB is consistent with relativity since it does not enable communication. This  device, as well as more general \emph{non-signalling} correlations have been studied extensively, mostly in terms of understanding the power and limitations of \emph{non-signalling} theories\cite{BLM+05,BM06,BBL+06,BS09}, as well as more generally in terms of \emph{information causality}\cite{PPK+09,ABPS09,CSS10} and \emph{local orthogonality}\cite{FSA+13,SFA+14}; see also\cite{NW09,NGHA15}.
One striking consequence of the NLB is that it implies the \emph{collapse} of classical communication complexity\cite{vD13}, meaning that,  any Boolean function can be computed in a two-party distributed context with \emph{a single bit of communication}, as long as the parties have access to the NLB correlations\footnote{This result was also shown by Richard Cleve (unpublished).}. This is presented as evidence against physical theories that allows the strong correlations of the~NLB.

Here, we make progress towards the question of secure position-based quantum cryptography by showing an efficient attack to \emph{any} scheme, where the participants are allowed the additional NLB resource.
 Our technique consists in showing that instantaneous nonlocal quantum computation is possible
 with a \emph{linear} amount of pre-shared entanglement (in the size of the circuit), together with a linear amount of uses of the NLB.
Furthermore, if we restrict the output to being a single qubit (say, held by Alice), the classical communication reduces to only two bits sent from Bob to Alice (in the case of quantum output), or a single bit (in the case of classical output). In both cases, this is optimal\cite{BBC+93}.  Thus our construction establishes a quantum analogue of the classical communication complexity collapse\cite{vD13} under no-signalling correlations.

\emph{Construction.\textbf{---}}Our construction builds on the techniques of teleportation\cite{BBC+93}, gate teleportation\cite{GC99}, and quantum computing on encrypted data\cite{Chi05,DNS10,FBS+14,Bro15,BJ15,BFK09} (see also  \cite{ZLC00,CLN05}).
A key observation is that the Pauli-$\xgate$ and~$\zgate$ corrections used in teleportation correspond precisely to the process of quantum one-time pad encryption\cite{AMTW00}.
Thus, we  view the two-party computation as being evaluated on encrypted quantum data, where the classical keys are available via the teleportation corrections. More precisely,  for each wire~$i$ in the computation, Alice keeps track of encryption keys $x_i^A \in \{0,1\}$ and $z_i^A \{0,1\}$ (Bob does likewise with values $x_i^B \in \{0,1\}$ and $z_i^B \{0,1\}$). At any point in the computation, the keys are \emph{distributed}: applying the operation $\xgate^{x_i^A  \oplus x_i^B}\zgate^{z_i^A  \oplus z_i^B}$ at each wire~$i$  results in the quantum state at that point in the (unencrypted) computation.
Crucially,  the parties can evaluate the circuit on encrypted data \emph{without any communication}: the  decryption being delayed until the end of the protocol, when the parties exchange the classical keys and thus can locally decrypt (reconstruct) their outputs\footnote{Inspired by a 2011 preliminary report on this work, Speelman\cite{Spe15arXiv}  used a similar framework to achieve instantaneous nonlocal quantum computation for circuits of low $\tgate$-depth; recently, these techniques have led to the breakthrough result of \emph{quantum fully homomorphic encryption}\cite{DSS16arXiv}}.

We represent the computation in the universal gateset 
 $\xgate: \ket{j}
  \mapsto \ket{j \oplus 1}$ and $\zgate: \ket{j} \mapsto (-1)^j\ket{j}$,  $\hgate : \ket{j} \mapsto
  \frac{1}{\sqrt{2}} (\ket{0} + (-1)^j \ket{1}$), $\pgate: \ket{j} \mapsto i^j\ket{j}$. 
  $\cnot: \ket{j}\ket{k} \mapsto
\ket{j}\ket{j \oplus k}$, $\rgate \ket{j} \mapsto e^{ij\pi/4}\ket{j}$, with all measurements being in the computational basis.
At the onset of the computation, Bob uses shared entanglement to teleport his input registers to Alice; instead of sending the required Pauli corrections, he updates his local keys to represent these correction values. For the input wires originally held by Alice, Bob sets the keys to~$0$. Alice sets all of her keys to~$0$. Next, Alice locally performs the computation. All Clifford gates ($\xgate$, $\zgate$, $\pgate$, $\hgate$, $\cnot$) are performed directly on the encrypted data, with both parties updating their keys after these gates, according to the well-known  relationships between Pauli matrices and Clifford group operations\cite{Got98} (see, \emph{e.g.}\cite{Chi05,DNS10,FBS+14,Bro15}).

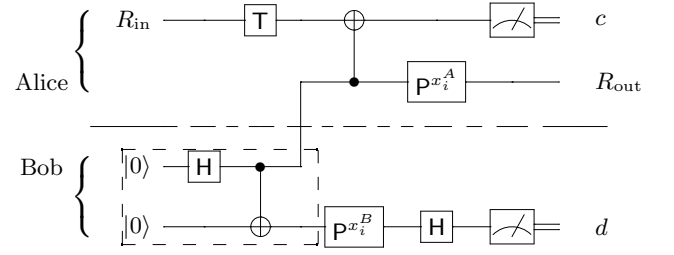
\begin{figure}[t!]
\centerline{
 \Qcircuit @C=1em @R=1em  {
&&&&&&\lstick{R_\text{in}} & \qw & \gate{\rgate} & \qw & \targ &\qw    &  \meter  &\cw &  \rstick{c}      \\
&\mbox{Alice}&&&& & & && & \ctrl{-1} &\gate{\pgate^{x_i^A}}    &  \qw &
\qw& \rstick{R_\text{out}}\\
&&&&\dw &\dw&\dw& \dw & \dw & \dw &\dw & \dw
 &  \dw & \dw  & \dw &\dw&\\
&\mbox{Bob}&& &&&\lstick{\ket{0}}    & \gate{\hgate}
&\ctrl{1} & \qw \qwx[-2]
 &   & & & & &&\\
&& &&&& \lstick{\ket{0}}    & \qw & \targ & \qw
 & \gate{\pgate^{x_i^B}}  & \gate{\hgate} &  \meter &  \cw
   & \rstick{d}
   \gategroup{4}{6}{5}{10}{1.4em}{--}
    \gategroup{1}{4}{2}{4}{0.7em}{\{}
 \gategroup{4}{4}{5}{4}{0.7em}{\{}
 }
 }
  \caption{\label{fig:R-gate-EPR}Entanglement-based protocol for a $\rgate$-gate on an input wire~$i$ held by Alice.
The input wire $R_{\text{in}} = \xgate^{x_i^A \oplus x_i^B} \zgate^{z_i^A \oplus z_i^B}\ket{\psi}$, and the output wire $R_{\text{out}} = \xgate^{x_i^A \oplus x_i^B \oplus c} \zgate^{x_i^A \oplus x_i^B \oplus z_i^A \oplus z_i^B \oplus x_i^A \cdot c \oplus (x_i^A \oplus c) \cdot x_i^B \oplus d}
\rgate \ket{\psi}$\,.
The circuit in
the dashed box prepares a two-qubit maximally entangled state and is executed before the computation begins.}
\end{figure}

The only remaining gate is the $\rgate$-gate. Although this is a single-qubit gate, it is not in the Clifford group, and thus does not allow a simple re-interpretation of the encryption key; in fact: $\rgate \xgate^a \zgate^b  = \xgate^a \zgate^{a \oplus b} \pgate^a \rgate$ (up to global phase). Various methods have been proposed to evaluate the $\tgate$ on encrypted data\cite{Chi05,DNS10,FBS+14,Bro15,BJ15}. We present in Fig.~\ref{fig:R-gate-EPR} a new method, that uses shared entanglement.   The encryption of the output includes a distributed \emph{multiplication}, $(x_i^A \oplus c) \cdot x_i^B$. Using the NLB correlations this can be re-linearized as $z^A \oplus z^B =  (x_i^A \oplus c) \cdot x_i^B$. The local key updates are therefore ${x'}_i^A = x_i^A  \oplus c$,  ${z'}_i^A =  z^A  \oplus x_i^A \oplus z_i^A \oplus x_i^A \cdot c   $, ${x'}_i^B = x_i^B $ and ${z'}_i^B =z^B \oplus  x_i^B  \oplus z_i^B \oplus d$.
Correctness of Figure~\ref{fig:R-gate-EPR} can be seen by quantum circuit manipulations and identities, as presented further on. We note that our construction shows that the $\tgate$-gate can be computed in the two-party setting without any communication (but with the use of an NLB). This improves on prior work that required quantum\cite{Chi05,DNS10} or classical\cite{BFK09} communication.

It remains to show that the  joint output of the computation can be obtained by a single round of simultaneous communication. This is accomplished by Alice using shared entanglement to teleport Bob's output registers to him; she then updates her Pauli keys accordingly. Next, both parties simultaneously exchange the classical keys required for decryption; a simple XOR calculation then allows each party to locally decrypt (reconstruct) their outputs
\footnote{We note that a variant of the protocol would forego the teleportation at both the beginning and the end of the protocol, instead using the \emph{nonlocal $\cnot$} procedure from\cite{DNS10}. The resulting protocol has essentially the same properties, but may be beneficial in certain circumstances.}.

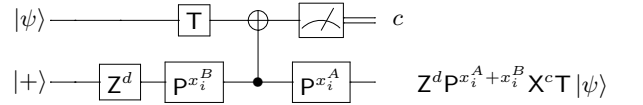
\begin{figure}[h!]
 \centerline{
 \Qcircuit @C=1em @R=1em {
&\hspace{-1.25cm}\mbox{$\ket{\psi}$} \qw & \qw&\gate{\tgate} &\targ      &  \meter & \cw   &  \rstick{\hspace{-.25cm} c}  &     \\
&\hspace{-1.25cm}\mbox{$\ket{+}$}   \qw & \gate{\zgate^d}&\gate{\pgate^{x_i^B} } &\ctrl{-1}  &   \gate{\pgate^{x_i^A}}   & \qw &
&\rstick{\hspace{-.25cm} \zgate^d \pgate^{x_i^A +x_i^B} \xgate^c \rgate\ket{\psi}}& &  }
 }
 \caption{Modified $\xgate$-teleportation circuit}
 \label{fig:correctness-T}
\end{figure}


\emph{Correctness of the $\tgate$-gate protocol.\textbf{---}}In order to show correctness of Fig.~\ref{fig:R-gate-EPR}, we consider a modification of the $\xgate$-teleportation circuit\cite{ZLC00} (Fig.~\ref{fig:correctness-T}), which can easily be seen as correct, since the diagonal gates $\zgate$ and $\pgate$ commute with control. Furthermore, on input $\xgate^{x_i^A \oplus x_i^B} \zgate^{z_i^A \oplus z_i^B}\ket{\psi}$,  
Fig.~\ref{fig:correctness-T} produces the same output as in Fig.~\ref{fig:R-gate-EPR}. Using the following identities (which hold up to global phase): $\pgate^{a \oplus b} = \zgate^{a \cdot b}\pgate^{a +b }$, $\tgate\xgate = \pgate\xgate\tgate$, $\tgate\zgate = \zgate\tgate$, $\xgate\zgate = \zgate\xgate$, $\pgate\xgate = \xgate\zgate\pgate$, $\pgate^2=\zgate$, we can compute the output as:

\begin{align*}
&\zgate^d \pgate^{x_i^A +x_i^B}  \xgate^c   \tgate \xgate^{x_i^A \oplus x_i^B} \zgate^{z_i^A \oplus z_i^B}\ket{\psi}\\
&= \zgate^{d\oplus  x_i^A \cdot x_i^B} \pgate^{x_i^A \oplus x_i^B}     \xgate^{c} \pgate^{x_i^A \oplus x_i^B}
\xgate^{x_i^A \oplus x_i^B } \zgate^{z_i^A \oplus z_i^B }  \tgate \ket{\psi}\\
&= \zgate^{d\oplus  x_i^A \cdot x_i^B} \xgate^{c} \zgate^{c \cdot (x_i^A \oplus x_i^B)} \zgate^{x_i^A \oplus x_i^B}     \xgate^{x_i^A \oplus x_i^B } \zgate^{z_i^A \oplus z_i^B}   \tgate \ket{\psi}\\
&= \xgate^{x_i^A \oplus x_i^B \oplus c}  \zgate^{  x_i^A \oplus x_i^B \oplus z_i^A \oplus z_i^B \oplus x_i^A \cdot c \oplus (x_i^A \oplus c) \cdot x_i^B \oplus d}  \tgate \ket{\psi}\\
\end{align*}

\emph{Consequences.\textbf{---}}
The impossibility of position-based quantum cryptography using nonlocal correlations follows as a  direct consequence of our construction.
As for the quantum analogue of the collapse of communication complexity, this follows by restricting the output to a single qubit (or bit) for Alice (and no output for Bob). In this case, Alice can reconstruct the output given only two classical bits from Bob (in the case that the output is classical, this is reduced to a single bit). This is optimal: in the quantum case, this follows from the optimality of teleportation\cite{BBC+93}, while in the classical case, any protocol with less than~1 bit of communication would violate relativity.

Since our result shows that  communication is not required for efficient instantaneous nonlocal quantum computation, we have established a no-go result for position-based quantum cryptography against efficient adversaries with non-signalling correlations. This implies that, if position-based quantum cryptography is indeed possible against efficient quantum adversaries, it will be thanks in part to bounds such as Tsirelson's\cite{Cir80}, according to which quantum mechanics is not maximally non-signalling.  One open question that remains is to characterize more broadly the set of physical theories that rule out position-based cryptography, for instance, in terms of non-signalling correlations that are not known to be distillable to the NLB, or other related  theories.
 \looseness=-1

\vspace{.25cm}

I would like to thank Gilles Brassard and Florian Speelman for fruitful discussions related to this work. This research is supported in part by Canada's \textsc{Nserc}.




\bibliographystyle{arxiv_no_month}

\end{document}